\documentclass[aps,prb,reprint,nolongbibliography,noeprint,onecolumn]{revtex4-2}
\usepackage{amsfonts,amssymb,amsmath,graphicx,color}
\usepackage[colorlinks=true,citecolor=blue,linkcolor=blue,urlcolor=blue]{hyperref}
\allowdisplaybreaks[4]
\newcommand\dc[1]{\dot {\cal #1}}
\newcommand\tc[1]{\tilde {\cal #1}}
\begin{document}
\title{Geometric spin-orbit coupling and chirality-induced spin selectivity}
\author{Atsuo Shitade}
\affiliation{Institute for Molecular Science, Aichi 444-8585, Japan}
\author{Emi Minamitani}
\affiliation{Institute for Molecular Science, Aichi 444-8585, Japan}
\date{\today}
\begin{abstract}

  We report a new type of spin-orbit coupling (SOC) called geometric SOC.
  Starting from the relativistic theory in curved space, we derive an effective nonrelativistic Hamiltonian in a generic curve embedded into flat three dimensions.
  The geometric SOC is $O(m^{-1})$, in which $m$ is the electron mass, and hence much larger than the conventional SOC of $O(m^{-2})$.
  The energy scale is estimated to be a hundred meV for a nanoscale helix.
  We calculate the current-induced spin polarization in a coupled-helix model as a representative of the chirality-induced spin selectivity.
  We find that it depends on the chirality of the helix and is of the order of $0.01 \hbar$ per ${\rm nm}$ when a charge current of $1~{\rm \mu A}$ is applied.
\end{abstract}
\maketitle
\section{Introduction} \label{sec:introduction}
Spin-orbit coupling (SOC) is a relativistic interaction between the spin and the orbital motion of an electron.
It gives rise to many intriguing phenomena such as the spin Hall~\cite{RevModPhys.87.1213} and the Edelstein effects~\cite{Ivchenko1978,Ivchenko1989,Aronov1989,Edelstein1990233}.
The spin Hall effect is a phenomenon in which the spin current flows perpendicular to an applied electric field, leading to the spin accumulation at the boundaries.
Its theoretical rediscovery~\cite{Murakami1348,PhysRevLett.92.126603} motivated the recent development in topological insulators~\cite{RevModPhys.82.3045,RevModPhys.83.1057}.
On the other hand, the Edelstein effect is a phenomenon in which spin polarization is induced by the electric field only when the inversion symmetry is broken.
These phenomena are governed by the energy scale of the SOC, which is proportional to $Z^4$ in atomic limit, $Z$ being the atomic number.
Thus, heavy elements are more likely to demonstrate nontrivial effects caused by the SOC.
In fact, the large spin Hall effect was observed in heavy metals
including Pt~\cite{PhysRevLett.98.156601,*PhysRevLett.98.249901,PhysRevLett.104.046601,PhysRevB.82.214403,PhysRevB.83.174405} and Au~\cite{Seki2008,PhysRevLett.104.046601,PhysRevB.82.214403},
and the Edelstein effect was observed in a Bi/Ag interface~\cite{Sanchez2013} and topological insulator surfaces~\cite{PhysRevLett.113.196601,nl502546c}.

In contrast, a spin filtering effect that resembles the Edelstein effect was reported
in chiral molecules composed of light elements~\cite{Ray814,Carmeli2002,PhysRevLett.96.036101,Gohler894,Mishra14872}.
The spin polarization of photoelectrons transmitted through the molecules depends on the molecular chirality.
Recently, the effect of chirality on the magnetoresistance~\cite{nl2021637,*nl2042062,Lueaay0571,2001.00097}
and the emergence of the Shiba states in a conventional superconductor~\cite{acs.nanolett.9b01552} were reported.
These phenomena are called chirality-induced spin selectivity (CISS)~\cite{jz300793y,annurev-physchem-040214-121554}. 
Surprisingly, the energy scale of the SOC relevant to the CISS was experimentally estimated as hundreds of meV~\cite{nl2021637,*nl2042062}, which is unexpected in light elements.
For instance, the conventional SOC in graphene was estimated to be $42.2~{\rm \mu eV}$~\cite{PhysRevLett.122.046403}.
Previous theoretical explanations of the CISS relied on the existence of
a large SOC~\cite{1.3167404,PhysRevB.85.081404,PhysRevLett.108.218102,Medina2012,PhysRevB.86.115441,jp401705x,1.4820907,PhysRevB.88.165409,Guo11658,PhysRevB.93.075407,acs.jpclett.9b02929}.
However, the origin of the SOC remains unclear.
Note that recently the orbital degree of freedom is recognized as another ingredient for the CISS~\cite{PhysRevB.102.035445,2008.08881}.

The CISS strongly indicates the existence of a large unknown SOC in chiral molecules.
The conventional SOC is derived from the Dirac Lagrangian density in electromagnetic field in flat spacetime.
Furthermore, a novel coupling between the spin and mechanical rotation was derived
from relativistic quantum mechanics in curved spacetime~\cite{PhysRevD.42.2045,PhysRevLett.106.076601,PhysRevB.84.104410}.
Since the chiral molecules are modeled as a one-dimensional ($1$D) curve embedded in $3$D flat space,
we can assume that the large SOC in chiral molecules originates from the relativistic effect in the $1$D curve.

In this paper, we derive an effective nonrelativistic Hamiltonian in a generic curve from the Dirac Lagrangian density in curved space.
We use the Frenet-Serret (FS) frame to describe the curve embedded in $3$D flat space~\cite{PhysRevA.23.1982,JPSJ.61.3825,jp401705x,PhysRevB.91.245412},
apply the thin-layer quantization to derive an effective Lagrangian density in the curve~\cite{JENSEN1971586,PhysRevA.23.1982},
and then perform the Foldy-Wouthuysen (FW) transformation to take the nonrelativistic limit~\cite{PhysRev.78.29,PhysRev.87.688}.
We find what we call geometric SOC of $O(m^{-1})$, in contrast to the conventional one of $O(m^{-2})$, where $m$ is the electron mass.
The energy scale is estimated to be a hundred meV.
We also calculate the current-induced spin polarization in a coupled-helix model
and find that it is of the order of $0.01 \hbar$ per ${\rm nm}$ when a charge current of $1~{\rm \mu A}$ is applied.

\section{Derivation of the geometric SOC} \label{sec:derivation}
We begin with the Dirac Lagrangian density in curved spacetime~\cite{9780521877879},
\begin{equation}
  L
  = e {\bar \psi} [\hbar \gamma^a e_a^{\phantom{a} \mu} (\partial_{\mu} + {\rm i} {\bar \omega}_{ab \mu} \Sigma^{ab}/4) - m] \psi. \label{eq:frenet-serret-dirac1}
\end{equation}
$e^a_{\phantom{a} \mu}$ is a vielbein, which is related to a metric as $g_{\mu \nu} = \eta_{ab} e^a_{\phantom{a} \mu} e^b_{\phantom{b} \nu}$,
whereas $e_a^{\phantom{a} \mu}$ and $e \equiv \det e^a_{\phantom{a} \mu}$ are the inverse and the determinant of the vielbein, respectively.
${\bar \omega}_{ab \mu}$ is the torsion-free spin connection calculated as
\begin{equation}
  {\bar \omega}_{ab \mu}
  = (e_a^{\phantom{a} \nu} t_{b \mu \nu} - e_b^{\phantom{b} \nu} t_{a \mu \nu} - e_a^{\phantom{a} \rho} e_b^{\phantom{b} \sigma} e^c_{\phantom{c} \mu} t_{c \rho \sigma})/2, \label{eq:frenet-serret-dirac2}
\end{equation}
with $t_{a \mu \nu} \equiv \partial_{\mu} e_{a \nu} - \partial_{\nu} e_{a \mu}$.
$\gamma^a$ is the $\gamma$ matrix that satisfies $\{\gamma^a, \gamma^b\} = 2 \eta^{ab}$,
and $\Sigma^{ab} \equiv [\gamma^a, \gamma^b]/2 i$ is proportional to the spin.
We take the Minkowski metric as $\eta_{ab} = [-1, +1, +1, +1]$.
In this convention, the $\gamma$ matrices are expressed as $\gamma^{\hat 0} = {\rm i} \beta, \gamma^{\hat \imath} = {\rm i} \beta \alpha^{\hat \imath}$
with use of the Dirac matrices that satisfy $\{\alpha^{\hat \imath}, \alpha^{\hat \jmath}\} = 2 \eta^{{\hat \imath} {\hat \jmath}}, \{\alpha^{\hat \imath}, \beta\} = 0, \beta^2 = 1$.
$\psi$ is a four-component spinor, and ${\bar \psi} \equiv \psi^{\dag} \beta$ is the Dirac conjugate.

First, we define a coordinate system.
We introduce the FS frame to describe a generic curve ${\vec r}(s)$ parametrized by its arc length $s$.
The tangential, normal, and binormal vectors are defined as ${\vec T} \equiv {\vec r}^{\prime}, {\vec N} \equiv {\vec T}^{\prime}/\kappa, {\vec B} \equiv {\vec T} \times {\vec N}$
and satisfy the FS formula,
\begin{equation}
  \begin{bmatrix}
    {\vec T} \\
    {\vec N} \\
    {\vec B}
  \end{bmatrix}^{\prime}
  =
  \begin{bmatrix}
    0 & \kappa & 0 \\
    -\kappa & 0 & \tau \\
    0 & -\tau & 0
  \end{bmatrix}
  \begin{bmatrix}
    {\vec T} \\
    {\vec N} \\
    {\vec B}
  \end{bmatrix}. \label{eq:frenet-serret4}
\end{equation}
Here, $\kappa$ and $\tau$ are the curvature and the torsion, respectively.
Right- ($\chi = +1$) and left-handed ($\chi = -1$) helices are described by ${\vec r} = [R \cos s/{\tilde R}, R \sin s/{\tilde R}, \chi P s/{\tilde R}]$,
where $R$ and $2 \pi P$ are the radius and the pitch, respectively, and ${\tilde R} \equiv \sqrt{R^2 + P^2}$.
The FS vectors are expressed as
\begin{subequations} \begin{align}
  {\vec T}
  = & \frac{1}{\tilde R}
  \begin{bmatrix}
    -R \sin s/{\tilde R} \\
    R \cos s/{\tilde R} \\
    \chi P
  \end{bmatrix}, \label{eq:frenet-serret5a} \\
  {\vec N}
  = &
  \begin{bmatrix}
    -\cos s/{\tilde R} \\
    -\sin s/{\tilde R} \\
    0
  \end{bmatrix}, \label{eq:frenet-serret5b} \\
  {\vec B}
  = & \frac{1}{\tilde R}
  \begin{bmatrix}
    \chi P \sin s/{\tilde R} \\
    -\chi P \cos s/{\tilde R} \\
    R
  \end{bmatrix}. \label{eq:frenet-serret5c}
\end{align} \label{eq:frenet-serret5}\end{subequations}
The curvature and the torsion are $\kappa = R/{\tilde R}^2$ and $\tau = \chi P/{\tilde R}^2$, respectively.

\begin{figure}
  \centering
  \includegraphics[clip,width=0.25\textwidth]{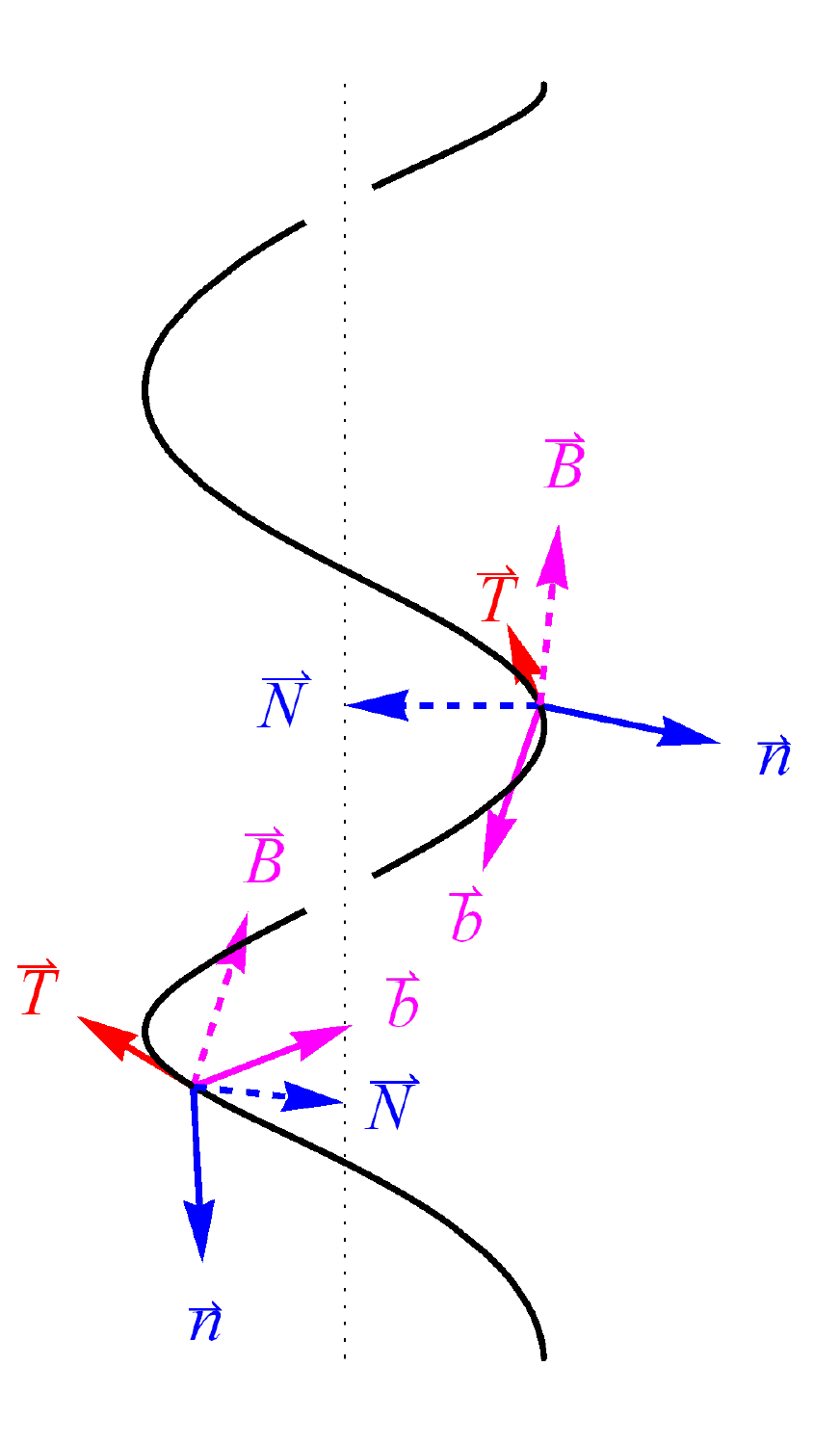}
  \caption{%
  The original (dashed) and the rotated (solid) FS frames in a right-handed helix.
  The tangential (${\vec T} = {\vec t}$), normal (${\vec N}, {\vec n}$), and binormal vectors (${\vec B}, {\vec b}$) are represented by red, blue, and magenta arrows, respectively.%
  } \label{fig:helix}
\end{figure}

We now shift to a rotated FS frame $\{{\vec t}, {\vec n}, {\vec b}\}$ following references~\cite{PhysRevA.23.1982,JPSJ.61.3825,jp401705x}.
These vectors are defined as
\begin{equation}
  \begin{bmatrix}
    {\vec t} \\
    {\vec n} \\
    {\vec b}
  \end{bmatrix}
  =
  \begin{bmatrix}
    1 & 0 & 0 \\
    0 & \cos \theta & -\sin \theta \\
    0 & \sin \theta & \cos \theta
  \end{bmatrix}
  \begin{bmatrix}
    {\vec T} \\
    {\vec N} \\
    {\vec B}
  \end{bmatrix}, \label{eq:frenet-serret6}
\end{equation}
in which $\theta$ is related to the torsion as
\begin{equation}
  \theta(s)
  \equiv \int_0^s {\rm d} s^{\prime} \tau(s^{\prime}). \label{eq:frenet-serret7}
\end{equation}
From equation~\eqref{eq:frenet-serret4}, we obtain
\begin{align}
  \begin{bmatrix}
    {\vec t} \\
    {\vec n} \\
    {\vec b}
  \end{bmatrix}^{\prime}
  = & \kappa
  \begin{bmatrix}
    0 & \cos \theta & \sin \theta \\
    -\cos \theta & 0 & 0 \\
    -\sin \theta & 0 & 0
  \end{bmatrix}
  \begin{bmatrix}
    {\vec t} \\
    {\vec n} \\
    {\vec b}
  \end{bmatrix}. \label{eq:frenet-serret8}
\end{align}
The original and the rotated FS frames in a right-handed helix are depicted in figure~\ref{fig:helix}.

We define a coordinate system for this frame as ${\vec x}(s, q_2, q_3) = {\vec r}(s) + {\vec n}(s) q_2 + {\vec b}(s) q_3$.
$q_2 = q_3 = 0$ describes the curve.
We obtain ${\rm d} {\vec x} = {\vec t} e {\rm d} s + {\vec n} {\rm d} q_2 + {\vec b} {\rm d} q_3$
with $e = 1 - \kappa q_2 \cos \theta - \kappa q_3 \sin \theta$ being the determinant of a vielbein chosen below.
The rotated FS frame is better than the original one because its metric is diagonal.
We choose a vielbein as
$e^{\hat 0}_{\phantom{\hat 0} 0} = 1,
e^{\hat \imath}_{\phantom{\hat \imath} s} = t^{\hat \imath} e,
e^{\hat \imath}_{\phantom{\hat \imath} q_2} = n^{\hat \imath},
e^{\hat \imath}_{\phantom{\hat \imath} q_3} = b^{\hat \imath}$
and
$e_{\hat 0}^{\phantom{\hat 0} 0} = 1,
e_{\hat \imath}^{\phantom{\hat \imath} s} = t_{\hat \imath}/e,
e_{\hat \imath}^{\phantom{\hat \imath} q_2} = n_{\hat \imath},
e_{\hat \imath}^{\phantom{\hat \imath} q_3} = b_{\hat \imath}$.
With this choice, $t_{a \mu \nu}$ defined above and the torsion-free spin connection equation~\eqref{eq:frenet-serret-dirac2} vanish.

The Dirac Lagrangian density equation~\eqref{eq:frenet-serret-dirac1} is expressed as
\begin{equation}
  L
  = \psi^{\dag} (e {\rm i} \hbar \partial_0 + {\rm i} \hbar {\vec \alpha} \cdot {\vec t} \partial_s + e {\rm i} \hbar {\vec \alpha} \cdot {\vec n} \partial_{q_2}
  + e {\rm i} \hbar {\vec \alpha} \cdot {\vec b} \partial_{q_3} - e m \beta) \psi. \label{eq:frenet-serret-dirac3}
\end{equation}
Rescaling the wave function as $\psi = e^{-1/2} \psi^{(0)}$, we obtain
\begin{subequations} \begin{align}
  L
  = & \psi^{(0) \dag} ({\rm i} \hbar \partial_0 - m \beta - {\cal O}^{(0)}) \psi^{(0)}, \label{eq:frenet-serret-dirac4a} \\
  {\cal O}^{(0)}
  = & -{\rm i} \hbar {\vec \alpha} \cdot [({\vec t} {\tilde \partial}_s + {\vec t}^{\prime}/2 e) + {\vec n} \partial_{q_2} + {\vec b} \partial_{q_3}], \label{eq:frenet-serret-dirac4b}
\end{align} \label{eq:frenet-serret-dirac4}\end{subequations}
with ${\tilde \partial}_s \equiv e^{-1/2} \partial_s e^{-1/2}$.
This Lagrangian density is obviously Hermitian.
Below, we use the Dirac representation ${\vec \alpha} = {\vec \sigma} \otimes {\tilde \sigma}^1, \beta = 1 \otimes {\tilde \sigma}^3, {\vec \Sigma} = {\vec \sigma} \otimes 1$,
in which $\sigma^{\hat \imath}, {\tilde \sigma}^i$ are the Pauli matrices for the spin and particle-hole degrees of freedom, respectively.
In equation~\eqref{eq:frenet-serret-dirac4a}, the first two terms are diagonal,
while ${\cal O}^{(0)}$ is off-diagonal and anticommutes with $\beta$.

Applying the thin-layer quantization, we derive an effective Lagrangian density in the curve~\cite{JENSEN1971586,PhysRevA.23.1982}.
We introduce a strong confinement potential in the normal and binormal directions.
This enables us to assume a separable wave function and integrate equation~\eqref{eq:frenet-serret-dirac4} with respect $q_2, q_3$.
We obtain an effective Lagrangian density for the tangential part $\psi_t^{(0)}$,
\begin{subequations} \begin{align}
  L_t
  = & \psi_t^{(0) \dag} ({\rm i} \hbar \partial_0 - m \beta - {\cal O}_t^{(0)}) \psi_t^{(0)}, \label{eq:frenet-serret-dirac5a} \\
  {\cal O}_t^{(0)}
  = & -{\rm i} \hbar {\vec \alpha} \cdot ({\vec t} \partial_s + {\vec t}^{\prime}/2). \label{eq:frenet-serret-dirac5b}
\end{align} \label{eq:frenet-serret-dirac5}\end{subequations}
Carrying out the FW transformation to take the nonrelativistic limit~\cite{PhysRev.78.29,PhysRev.87.688},
we find that under a unitary transformation $\psi_t^{(0)} = e^{-\beta O_t^{(0)}/2 m} \psi_t^{(1)}$, the Lagrangian density can be approximated as
\begin{equation}
  L_t
  = \psi_t^{(1) \dag} [{\rm i} \hbar \partial_0 - m \beta - \beta ({\cal O}_t^{(0) 2} + {\rm i} \hbar {\dc O}_t^{(0)})/2 m] \psi_t^{(1)}, \label{eq:frenet-serret-dirac6}
\end{equation}
up to $O(m^{-1})$.
${\cal O}_t^{(0) 2}/2 m$ is diagonal, while ${\rm i} \hbar {\dc O}_t^{(0)}/2 m$ is off-diagonal.
Retaining the upper block in the diagonal terms and dropping the rest mass energy,
we obtain the effective nonrelativistic Hamiltonian for a two-component spinor $\chi_t^{(1)}$,
\begin{equation}
  {\cal H}_t^{(1)}
  = p_s^2/2 m + \hbar^2 \kappa^2/8 m + (\hbar/2 m) \{p_s, \kappa {\vec \sigma} \cdot {\vec B}\}/2
  = (p_s + \hbar \kappa {\vec \sigma} \cdot {\vec B}/2)^2/2 m. \label{eq:frenet-serret-dirac7}
\end{equation}
The third term includes the momentum $p_s \equiv -{\rm i} \hbar \partial_s$ and the spin ${\vec \sigma} \cdot {\vec B}$,
in which ${\vec B}$ is the binormal vector in the original FS frame.
We term this as the geometric SOC.
The second term enables completing the square,
and the form of $p_s + \hbar \kappa {\vec \sigma} \cdot {\vec B}/2$ reflects the relativistic nature characterized by the conservation of the total angular momentum,
as demonstrated later.

\section{Notes on the geometric SOC} \label{sec:notes}
First, we emphasize that the geometric SOC is completely different from the conventional SOC
because the former is $O(m^{-1})$, while the latter is $O(m^{-2})$.
A similar but non-Hermitian result was reported in reference~\cite{JPSJ.61.3825}.
We believe that our result is physically correct, since we started with the Hermitian Lagrangian density and performed the unitary transformation.
It was reported that the uniform accelaration, denoted by ${\vec a}$, generates another SOC of $O(m^{-1})$,
${\cal H}_a = (\hbar/4 m) {\vec \sigma} \cdot {\vec a} \times {\vec p}$~\cite{PhysRevD.42.2045}.
If we assume ${\vec a}/c^2 = -2 \kappa {\vec N}$ owing to the confinement potential, in which we have restored the speed of light $c$,
${\vec p}$ is parallel to ${\vec T}$, and the form becomes the same as that of the geometric SOC.
Although this correspondence is useful for the intuitive picture of the geometric SOC,
its validity is not obvious because $\kappa$ and ${\vec N}$ depend on the arc length.
Moreover, in contrast to the acceleration-induced SOC, the geometric SOC is intrinsic in $1$D curves.

The geometric SOC is of the same form as the SOC that has been assumed
in the theoretical literature~\cite{PhysRevB.85.081404,PhysRevLett.108.218102,PhysRevB.86.115441,1.4820907,Guo11658,PhysRevB.93.075407}.
In the case of DNA, the radius and the pitch are $R = 1~{\rm nm}$ and $2 \pi P = 3.2~{\rm nm}$, respectively~\cite{jz300793y}, leading to the curvature $\kappa = 0.8~{\rm nm}^{-1}$.
The energy scale which is estimated to be $\hbar v_{\rm F} \kappa/2 = 160~{\rm meV}$, using a typical Fermi velocity $v_{\rm F} = 6 \times 10^5~{\rm m/s}$~\cite{annurev-physchem-040214-121554},
is of the same order of magnitude as the experimental result~\cite{nl2021637,*nl2042062}.
In addition, the obtained geometric SOC is consistent with the experimental fact that the CISS was not observed in a single-stranded DNA but in a double-stranded one~\cite{Gohler894}.
As already pointed out~\cite{PhysRevLett.108.218102},
any SOC of the first order with respect to the momentum can be eliminated by a unitary transformation
when the system is a $1$D curve without any sublattice degree of freedom.
This is also true for the geometric SOC.
In fact, we obtain ${\tc H}_t^{(1)} = p_s^2/2 m$ by the following unitary transformation,
\begin{equation}
  \chi_t^{(1)}(s)
  = {\rm P} \exp \left[-\frac{\rm i}{2} \int_0^s {\rm d} s^{\prime} \kappa(s^{\prime}) {\vec \sigma} \cdot {\vec B}(s^{\prime})\right] {\tilde \chi}_t^{(1)}(s). \label{eq:ciss1}
\end{equation}
Here, ${\rm P}$ is the path-ordered product along $s$.
Therefore, the sublattice degree of freedom is essential for spin-related phenomena.
In the case of single-stranded $\alpha$-helical protein, the long-range SOC plays the role~\cite{Guo11658},
and indeed the CISS was experimentally observed~\cite{Mishra14872}.

We may carry out the FW transformation before the thin-layer quantization, but this leads to a different result,
\begin{equation}
  {\cal H}_t^{(1)}
  = \frac{p_s^2}{2 m} - \frac{\hbar^2 \kappa^2}{8 m}, \label{eq:frenet-serret-dirac8}
\end{equation}
The second term is the quantum geometric potential owing to the curvature~\cite{JENSEN1971586,PhysRevA.23.1982}.
The geometric SOC does not appear.
Similar noncommutativity has already been pointed out in the context of a curved surface~\cite{PhysRevA.48.1861,BRANDT20163036}.
This problem may originate from the above assumption of the separability
and may be resolved by a similar approach based on the projection to the wave function of the normal part of the nonrelativistic kinetic term~\cite{acs.jpcc.9b05020,Geyer2020}.
However, it seems difficult to apply this approach directly to the relativistic theory and remains to be resolved.

The relevance of the geometric SOC becomes evident when we consider a $1$D ring.
In this case, $s = R \phi, \kappa = R^{-1}, \tau = 0, {\vec B} = {\vec z}$, and equation~\eqref{eq:frenet-serret-dirac7} is reduced to
\begin{equation}
  {\cal H}_t^{(1)}
  = \frac{\hbar^2}{2 m R^2}(-{\rm i} \partial_{\phi} + \sigma^{\hat 3}/2)^2. \label{eq:frenet-serret-dirac9}
\end{equation}
Since $\ell^{\hat 3} \equiv -{\rm i} \partial_{\phi}$ is the orbital angular momentum, $\ell^{\hat 3} + \sigma^{\hat 3}/2$ is the total one.
In the relativistic theory, only the total angular momentum is conserved.
When the subspace is curved, the same occurs even in the nonrelativistic limit.
In fact, the spectra obtained from equation~\eqref{eq:frenet-serret-dirac9} are consistent with the nonrelativistic limit of the Dirac spectra in the $1$D ring~\cite{Cotaescu_2007}.
Experimentally, such spectra have not been observed in quantum rings.
This is because the quantum rings are fabricated on semiconductor heterostructures and connected to $2$D electron gases, which are nonrelativistic.

\section{Edelstein effect caused by the geometric SOC} \label{sec:edelstein}
\begin{figure*}
  \centering
  \includegraphics[clip,width=0.98\textwidth]{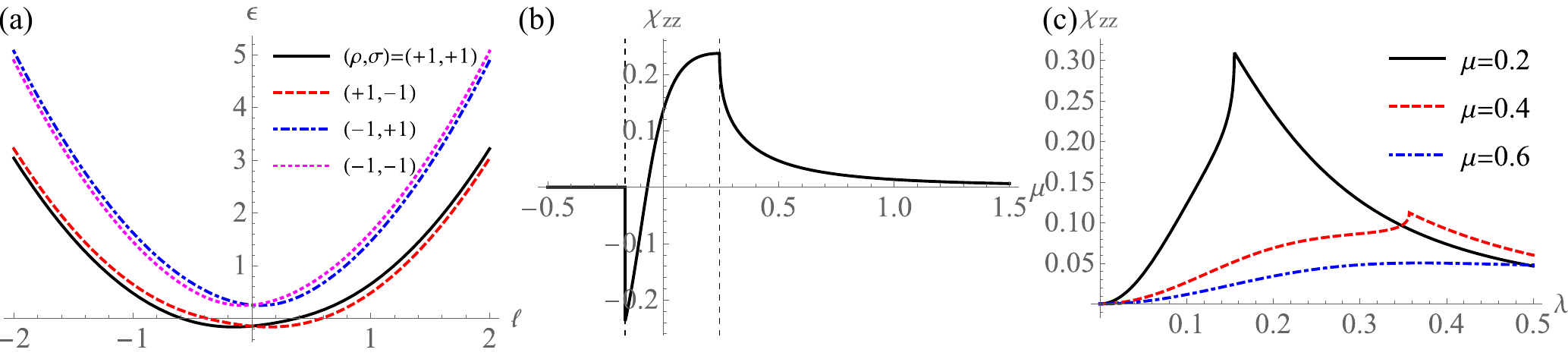}
  \caption{%
  (a) Band structure of the coupled-helix model.
  The black (solid) and the red (dashed) lines represent the two lower bands with $\rho = +1$,
  while the blue (dot-dashed) and the magenta (dotted) lines represent the two upper bands with $\rho = -1$.
  (b) The dimensionless Edelstein coefficient $\chi_{zz}$ for the interhelix coupling $\lambda = 0.2$ as a function of the chemical potential $\mu$.
  The dotted lines represent the band edges.
  (c) The $\lambda$ dependence of $\chi_{zz}$ for $\mu = 0.2, 0.4, 0.6$.
  We use the radius $R = 1~{\rm nm}$ and the pitch $2 \pi P = 3.2~{\rm nm}$.
  } \label{fig:ciss}
\end{figure*}

Finally, we calculate the current-induced spin polarization in the Edelstein effect~\cite{Ivchenko1978,Ivchenko1989,Aronov1989,Edelstein1990233}.
Since this phenomenon requires the presence of an SOC and the inversion symmetry breaking, it would be a sort of the CISS phenomena.
For the aforementioned reason, we consider a model in which ${\cal H}_t^{(1)}(s)$ and ${\cal H}_t^{(1)}(s + \pi {\tilde R})$ are coupled via a constant coupling $\Lambda$~\cite{PhysRevLett.108.218102}.
This model describes two coupled helices as in the double-stranded DNA~\cite{PhysRevLett.108.218102}.
According to equation~\eqref{eq:frenet-serret5c}, the ${\hat 1}, {\hat 2}$ components of ${\vec B}$ change their signs by $s \rightarrow s + \pi {\tilde R}$, while the ${\hat 3}$ component does not.
Thus, the Hamiltonian of the coupled-helix model is expressed as
\begin{equation}
  {\cal H}
  = p_s^2/2 m + \hbar^2 \kappa^2/8 m + \Lambda \rho^1 + B_{\rm Z} \sigma^{\hat 3}
  + (\hbar \kappa/2 m) (\sigma^{\hat 3} B_{\hat 3} p_s + \rho^3 \{p_s, {\vec \sigma}_{\perp} \cdot {\vec B}_{\perp}\}/2). \label{eq:ciss2}
\end{equation}
Here, we have added the Zeeman coupling $B_{\rm Z} \sigma^{\hat 3}$ only for calculating the expectation value of $\sigma^{\hat 3}$ for each band and $p_s$.
The $s$ dependence in ${\vec B}_{\perp}$ can be eliminated by a unitary transformation $\chi = e^{-{\rm i} (s/{\tilde R}) \sigma^{\hat 3}/2} {\tilde \chi}$,
which sends $p_s \rightarrow p_s - \hbar \sigma^{\hat 3}/2 {\tilde R}, {\vec \sigma}_{\perp} \cdot {\vec B}_{\perp} \rightarrow -(\chi P/{\tilde R}) \sigma^{\hat 2}$.
For convenience, we introduce the dimensionless parameters $p_s = \hbar \ell_s/{\tilde R}, R = {\tilde R} r, P = {\tilde R} p, \Lambda = E_0 \lambda, B_{\rm Z} = E_0 b_{\rm Z}$ with $r^2 + p^2 = 1, E_0 \equiv \hbar^2/2 m {\tilde R}^2$.
The dimensionless transformed Hamiltonian now takes the form
\begin{equation}
  {\tilde h}
  = \ell_s^2 + p^2/4 + \lambda \rho^1 + b_{\rm Z} \sigma^{\hat 3} - (p^2 \sigma^{\hat 3} + \chi r p \rho^3 \sigma^{\hat 2}) \ell_s, \label{eq:ciss4}
\end{equation}
and its eigenvalues are
\begin{subequations} \begin{align}
  \epsilon_{\rho \sigma \ell}
  = & \ell^2 + p^2/4 - \rho \delta_{\sigma \ell}, \label{eq:ciss5a} \\
  \delta_{\sigma \ell}
  \equiv & \sqrt{p^2 (\ell - b_{\rm Z} - \sigma \lambda)^2 + r^2 (b_{\rm Z} + \sigma \lambda)^2}, \label{eq:ciss5b}
\end{align} \label{eq:ciss5}\end{subequations}
for each $\rho = \pm 1, \sigma = \pm 1, \ell$,
from which we obtain the group velocity $\langle v^z \rangle_{\rho \sigma \ell} = (\chi P/\hbar) E_0 \partial_{\ell} \epsilon_{\rho \sigma \ell}$
and the spin $\langle \sigma^{\hat 3} \rangle_{\rho \sigma \ell} = \partial_{b_{\rm Z}} \epsilon_{\rho \sigma \ell}$.
All the quantities correspond to $b_{\rm Z} = 0$.
The band structure for $R = 1~{\rm nm}, 2 \pi P = 3.2~{\rm nm}$, and $\lambda = 0.2$ is shown in figure~\ref{fig:ciss}(a).

When we apply an electric field $E_z$ in the $z$ direction, the charge current $j^z$ and the spin $s_z$ are induced in the same direction.
We calculate the electric conductivity $\sigma^{zz}$ and the Edelstein coefficient $\alpha^z_{\phantom{z} z}$,
which characterize $j^z = \sigma^{zz} E_z, s_z = \alpha^z_{\phantom{z} z} E_z$, respectively.
In the helix, since $z = \chi P s/{\tilde R}$, the electric field reduces to $E_z (\chi P/{\tilde R})$ in terms of the arc length $s$.
Within the relaxation time approximation at zero temperature, we obtain
\begin{subequations} \begin{align}
  \sigma^{zz}
  = & \tau_{\rm re} q^2 \sum_{\rho \sigma} \int \frac{{\rm d} \ell}{2 \pi {\tilde R}}
  (\langle v^z \rangle_{\rho \sigma \ell})^2 \delta(E_0 (\epsilon_{\rho \sigma \ell} - \mu)) \notag \\
  = & \frac{\tau_{\rm re} q^2 p^2}{4 \pi m {\tilde R}} \sum_{\rho \sigma} \int {\rm d} \ell
  [2 \ell - \rho p^2 (\ell - \sigma \lambda)/\delta_{\sigma \ell}]^2
  \delta(\ell^2 + p^2/4 - \rho \delta_{\sigma \ell} - \mu), \label{eq:ciss7a} \\
  \alpha^z_{\phantom{z} z}
  = & \tau_{\rm re} q \frac{\hbar}{2} \sum_{\rho \sigma} \int \frac{{\rm d} \ell}{2 \pi {\tilde R}}
  \langle \sigma^{\hat 3} \rangle_{\rho \sigma \ell} \langle v^z \rangle_{\rho \sigma \ell}
  \delta(E_0 (\epsilon_{\rho \sigma \ell} - \mu)) \notag \\
  = & \frac{\chi \tau_{\rm re} q p}{4 \pi} \sum_{\rho \sigma} \int {\rm d} \ell
  \rho (p^2 \ell - \sigma \lambda)/\delta_{\sigma \ell}
  [2 \ell - \rho p^2 (\ell - \sigma \lambda)/\delta_{\sigma \ell}]
  \delta(\ell^2 + p^2/4 - \rho \delta_{\sigma \ell} - \mu), \label{eq:ciss7b}
\end{align} \label{eq:ciss7}\end{subequations}
in which $\tau_{\rm re}$ is the relaxation time, $q$ is the electron charge, and $E_0 \mu$ is the chemical potential.
$\alpha^z_{\phantom{z} z}$ changes its sign when the chirality changes.
In figure~\ref{fig:ciss}(b), we show the $\mu$ dependence of the dimensionless Edelstein coefficient
$\chi_{zz} = (\chi q p/ m {\tilde R}) \alpha^z_{\phantom{z} z}/\sigma^{zz} = (q \tau/m) \alpha^z_{\phantom{z} z}/\sigma^{zz}$,
which characterizes $s_z = (m/q \tau) \chi_{zz} j^z$.
Note that $\chi_{zz}$ is independent of $\chi, \tau_{\rm re}$.
When the chemical potential lies only in the two lower bands, whose quantum number is $\rho = +1$,
$\chi_{zz}$ changes from negative values to positive ones,
because the integrand in equation~\eqref{eq:ciss7b} is approximated as $-p^2 < 0$ for small $|\ell|$ and $2 p |\ell| > 0$ for large $|\ell|$.
The two upper bands characterized by $\rho = -1$ have a negative contribution to $\chi_{zz}$.
Thus, $\chi_{zz}$ attains its the maximum value at the edge of the two upper bands.
We also show the $\lambda$ dependence of $\chi_{zz}$ for $\mu = 0.2, 0.4, 0.6$ in figure~\ref{fig:ciss}(c).
$\chi_{zz}$ vanishes for $\lambda = 0$ since the geometric SOC can be eliminated by a unitary transformation, as mentioned above.
Each peak position corresponds to the parameter for which the band edge is equal to the chemical potential.
We find that $\chi_{zz}$ is of the order of $0.1$ in a wide range of the parameters $\lambda, \mu$.
Therefore, $s_z \simeq 0.01 \hbar$ can be observed per ${\rm nm}$ when a charge current of $1~{\rm \mu A}$ is applied.
This unit of the charge current is not large compared with the experimental setup in reference~\cite{2001.00097}.

\section{Summary} \label{sec:summary}
To summarize, we have derived the geometric SOC of $O(m^{-1})$ starting from the Dirac Lagrangian density in curved space,
then applying the thin-layer quantization~\cite{JENSEN1971586,PhysRevA.23.1982} and finally, taking the nonrelativistic limit~\cite{PhysRev.78.29,PhysRev.87.688}.
If the order is reversed, the geometric SOC does not appear.
The estimated energy scale is a hundred meV for a nanoscale helix, much larger than the conventional SOC expected in light elements.
We have also calculated the Edelstein coefficient in the coupled-helix model, which describes two coupled helices.
The current-induced spin polarization depends on the chirality and is of the order of $0.01 \hbar$ per ${\rm nm}$ when a charge current of $1~{\rm \mu A}$ is applied.
Although we have not considered the detailed compositions or the structures of chiral molecules,
we believe that the emergence of the geometric SOC is general and provides a theoretical foundation for the CISS.
\begin{acknowledgments}
  We thank D~Hirobe, G~Tatara, and A~Kato for introducing the CISS and Y~Yanase for discussions on the Edelstein effect.
  This work was supported by the Japan Society for the Promotion of Science KAKENHI (Grants No.~JP18K13508 and No.~JP18H03880).
\end{acknowledgments}
%
\end{document}